\documentclass[a4paper,11pt]{article}
\pdfoutput=1 

\usepackage{jcappub} 

\usepackage[T1]{fontenc} 
\usepackage{hyperref}
\usepackage[dvipsnames, table]{xcolor}

\usepackage{cleveref}

\usepackage{slashed}
\usepackage{mathtools}
\usepackage{amssymb}
\usepackage{physics}
\usepackage{soul}

\usepackage{parskip}

\usepackage{feynmp-auto}

\usepackage{tcolorbox}

\newcommand{\be}{\begin{equation}}
\newcommand{\ee}{\end{equation}}
\newcommand{\bea}{\begin{eqnarray}}
\newcommand{\eea}{\end{eqnarray}}
\definecolor{lightblue}{rgb}{0.8,0.85,1}

\title{Using $\Delta N_{\rm eff}$ to constrain preferred axion model dark matter}

\author[a, b]{Andrew Cheek,}
\author[a, b]{Ui Min}

\affiliation[a]{Tsung-Dao Lee Institute \& School of Physics and Astronomy, Shanghai Jiao Tong
University, Shanghai 200240, China}

\affiliation[b]{Key Laboratory for Particle Astrophysics and Cosmology (MOE) \& Shanghai Key Laboratory for Particle Physics and Cosmology, Shanghai Jiao Tong University, Shanghai 200240, China}

\emailAdd{acheek@sjtu.edu.cn}
\emailAdd{ui.min@sjtu.edu.cn}

\abstract{Preferred axion models are minimal realizations of the Peccei-Quinn solution to the strong CP problem while providing a dark matter candidate. These models invoke new heavy quarks that interact strongly with the Standard Model bringing them into thermal equilibrium in the early Universe. We show that for a number of these models, the heavy quarks will decay after axions have decoupled from the Standard Model thermal bath. As a consequence, any axion products in the decay form a component of dark radiation. This provides the potential to differentiate between preferred axion models through measurements of the number of relativistic degrees of freedom. The most sensitive of which comes from the Planck collaboration's measurements of the Cosmic Microwave Background. We find that existing constraints allow us to rule out regions of parameter space for 40\% of the canonical preferred axion models. }

\begin{document}
\maketitle
\flushbottom

\section{Introduction}

In the last century, the understanding of the constituents and dynamics of the Universe has reached great depths, but at the same time, posed profound challenges~\cite{Baumann:2022mni}. Discovering the fundamental nature of dark matter is one of these challenges~\cite{Cirelli:2024ssz}. Fortunately, some of the most elegant solutions to the dark matter problem provide testable predictions. Two such solutions are weakly interacting massive particle (WIMP) dark matter and axion dark matter. For WIMPs, a stable particle will undergo thermal freeze-out and obtain the correct relic abundance for masses around $1\,{\rm GeV}$ to $100\, {\rm TeV}$~\cite{Lee:1977ua, Griest:1989wd}. This mechanism was found concurrently with developments of the supersymmetry, evidence of which was expected to appear at similar scales. However, in the last decade, tremendous experimental achievements have put serious pressure on this scenario~\cite{ATLAS:2024lda, LZ:2022lsv, XENON:2023cxc, PandaX:2024qfu}. 

On the other hand, axion dark matter is a byproduct of a popular solution to the strong CP problem in the Standard Model (SM) of particle physics. The problem arises from the lack of observed CP violation in strong interactions despite the fact that a priori it is expected via the $\theta$ term. The Peccei-Quinn mechanism solves this problem by introducing a new global $U(1)_{\rm PQ}$ symmetry. This symmetry is then spontaneously broken resulting in a light pseudo-goldstone boson, the axion~\cite{Peccei:1977hh, Peccei:1977np,Weinberg:1977ma, Wilczek:1977pj}. This axion contributes to the CP-violating term dynamically. Under the influence of the QCD potential, the axion field is driven to a vacuum expectation value that precisely cancels the bare $\theta$ term. 

Interestingly, during the expansion of the Universe, this axion field undergoes a non-thermal production mechanism called the misalignment mechanism~\cite{Preskill:1982cy, Dine:1982ah, Abbott:1982af}. Here, a thermally decoupled scalar is generated through field oscillations around its vacuum expectation value. This enables one to identify the axion as a dark matter candidate. Much like WIMP dark matter, interactions with the SM are required for the picture to work and therefore, the theory is testable. Axion dark matter is currently the next major lamppost in which experimentalists will probe~\cite{Irastorza:2018dyq, Sikivie:2020zpn, Billard:2021uyg, Adams:2022pbo}.

The PQ solution holds greater phenomenological implications than a dark matter candidate. Indeed, attempts to make the solution more theoretically complete require additional particle content and topological defects~\cite{Zeldovich:1974uw, Sikivie:1982qv}. In general these additions can be inconsistent with the Universe we observe. To determine which axion models were free from such issues, authors of Refs.~\cite{DiLuzio:2016sbl, DiLuzio:2017pfr} introduced the \textit{preferred axion models}. In this paper we take a closer look at these models by following the evolution of the new particles, which are required to be unstable. Despite their instability, we find that their presence can alter the relative abundance of axions and photons at the time of recombination. This should be observable in measurements of the number of additional relativistic degrees of freedom, $\Delta N_{\rm eff}$. We find that only 40\% of such models are equivalent to the standard expectation from axion cosmology, and some models' parameter space is already excluded by existing measurements~\cite{Planck:2018jri}.

The contribution that QCD axions or axion-like particles may have on $\Delta N_{\rm eff}$ has been widely studied~\cite{Masso:2002np,Hannestad:2003ye, Melchiorri:2007cd,Graf:2010tv,Salvio:2013iaa, Ferreira:2018vjj,Arias-Aragon:2020shv, Giare:2020vzo, DEramo:2021lgb, Caloni:2022uya,Bianchini:2023ubu,Badziak:2024szg,Sakurai:2024cbi}. These works typically calculate the contribution from thermal processes. However, in this paper, we explore an implicit but often overlooked consequence of some of the preferred axion models. Specifically, the out-of-equilibrium injection of axions from heavy quark decay produces a highly boosted population of axions with respect to their thermal counterparts. Our results show that under the standard assumptions of the preferred axion model framework, the boosted axion component in the Universe can be large. 

This paper is structured in the following way: in section~\ref{sec:pref_ax} we review the preferred axion model framework and enumerate the models; in section~\ref{sec:decay_width} we write the leading decay terms for the new heavy particles for each model; we show our predictions for the cosmology of each model in section~\ref{sec:dNeff}, validating our approximations and finally comparing our results for $\Delta N_{\rm eff}$ to current bounds; we conclude in section~\ref{sec:conclusion}. 

\section{Preferred axion models and heavy Quark decays}
\label{sec:pref_ax}

The PQ mechanism relies on the introduction of a new $U(1)_{\rm PQ}$ global symmetry which is spontaneously broken at some scale $f_a$. The breaking gives way to the pseudo-goldstone boson, the axion~\cite{Weinberg:1977ma, Wilczek:1977pj}, which has the effective Lagrangian, 
\begin{equation}
\mathcal{L}_a = \frac{1}{2} (\partial_\mu a)^2 + \mathcal{L}(\partial_\mu a, \psi) + \frac{g_s^2}{32 \pi^2} \frac{a}{f_a} G\tilde{G},
\label{eq:effective_Lag}
\end{equation}
where the last term is the dimension 5 anomaly term required for solving the strong CP problem. This term must be generated by some high-scale extension to the SM of which there are many. By investigating both the field theoretic and cosmological effects of these extensions the authors of Refs.~\cite{DiLuzio:2016sbl, DiLuzio:2017pfr} proposed the \textit{preferred axion models}, which meet the following criteria:
\begin{enumerate}
    \item They require $N_{\text{DW}} = 1$ to avoid the domain wall problem.
    \item The axion decay constant $f_a$ must lie within the range $5 \times 10^9 - 3 \times 10^{11} \text{ GeV}$ to reproduce the observed dark matter abundance through the misalignment mechanism or cosmic string decay.\footnote{The upper bound on $f_a$ has been introduced by Ref.~\cite{Alonso-Alvarez:2023wig}, we elaborate more on this in the text.}
    \item Any new strongly interacting fermion, $Q$, must have gauge charges that allow it to couple to SM states with operators of dimension $\leq 5$.
    \item Models must not induce a Landau pole in the SM gauge couplings below the Planck-scale.   
\end{enumerate}
It is straightforward to see that the first criteria is only relevant for the case where $U(1)_{\rm PQ}$ is broken after the end of inflation when the reheating temperature is above the PQ breaking scale, $T_{\rm RH}> T_{\slashed{\rm PQ}}$. If breaking occurred before inflation and $T_{\rm RH}< T_{\slashed{\rm PQ}}$, topological defects such as domain walls would be inflated away, and $N_{\text{DW}}>1$ would no longer cause issues cosmologically~\cite{Guth:1980zm}. Therefore, the preferred axion model framework is concerned with the post-inflationary breaking scenario, which is the focus of this paper.

Insisting that stable domain walls are not formed and without providing a mechanism for their destruction~\cite{Barr:2014vva, Reig:2019vqh, Caputo:2019wsd} substantially restricts the axion models available. For example, models that imbue SM particles with $U(1)_{\rm PQ}$ charge, known as DFSZ models~\cite{Zhitnitsky:1980tq,Dine:1981rt}, typically have $N_{\rm DW}=3$ or $N_{\rm DW}=6$. It has been shown that flavour-dependent DFSZ models can achieve $N_{\rm DW}=1$~\cite{Bardeen:1978nq, Davidson:1983tp,Davidson:1984rr,Cox:2023squ}. This possibility is interesting but more streamlined solutions are found in KSVZ type models~\cite{Kim:1979if,Shifman:1979if}. In these models SM particles are uncharged under the $U(1)_{\rm PQ}$, and they introduce a new strongly coupled chiral fermion $Q$ which is charged under $U(1)_{\rm PQ}$. Within the KSVZ setup, Refs.~\cite{DiLuzio:2016sbl, DiLuzio:2017pfr} found that there are only two charge assignments that allow for $N_{\rm DW}=1$ and $d\leq 5$ decays, KSVZ-I and KSVZ-II, which have the $(SU(3)_c, SU(2)_L, U(1)_Y)$ charges $(3, 1, -1/3)$ and $(3, 1, +2/3)$, respectively. The relevant Lagrangian terms for the KSVZ type models are
\begin{equation}
 \mathcal{L}_{\rm PQ}= \vert\partial_\mu\Phi\vert^2 +\overline{Q}i\slashed{D}Q - (y_Q\overline{Q}_L Q_R \Phi + {\rm H.c})\,.
    \label{eq:LagPQ}
\end{equation}
where $\slashed{D}$ is the covariant derivative and $\Phi$ is the new complex scalar which is associated with the spontaneous $U(1)_{\rm PQ}$ breaking, it has a PQ charge $1$, and no charge under the SM gauge group. The axion field $a$ is obtained through symmetry breaking in the standard way
\begin{equation}
    \Phi\to\frac{1}{\sqrt{2}}\left(f_a+\phi\right)\,e^{ia/f_a},
    \label{eq:phi_broke}
\end{equation}
where $f_a$ is the aforementioned `decay constant' and $\phi$ is the radial mode. It is through this breaking that the Yukawa term leads to a massive $Q$ field with $m_Q=y_Q f_a/\sqrt{2}$. For the majority of this paper we take $y_Q=1$ and will comment explicitly on the effects of relaxing this choice. The axion, at this level is massless, however through its interaction with gluons, it obtains a mass below the QCD phase transition~\cite{Weinberg:1977ma}
\begin{equation}
    m_a = 5.7\times10^{-5}\,{\rm eV} \left(\frac{10^{11}\,{\rm GeV}}{f_a}\right).
    \label{eq:QCDaxionmass}
\end{equation}
Since both the mass and gluon interactions come from the same term, there is a clear predicted parameter space that is the target for experimental searches. In particular, the most promising coefficient to probe is the axion-photon coupling, $g_{a\gamma}$.

Astrophysical objects provide a powerful tool for constraining the axion mass window~\cite{Raffelt:1990yz, Raffelt:2006cw, Caputo:2024oqc}. The most constraining being the limits on axion emission from supernovae, which bound hadronic axion models to $f_a\gtrsim 1.4\times 10^8\,{\rm GeV}$~\cite{Carenza:2019pxu}. Additionally, the KSVZ type axion models induce flavour violating decays of hadrons, providing terrestrial tests from experiments such as NA62~\cite{NA62:2017rwk} and KOTO~\cite{Yamanaka:2012yma}. The effective bounds on $f_a$ can be taken from Ref.~\cite{MartinCamalich:2020dfe} and the specific results for KSVZ types I and II where considered in Ref.~\cite{Alonso-Alvarez:2023wig}. However, for the mass region most relevant for axion dark matter, around $10\,{\rm \mu eV}$, haloscope experiments are most promising~\cite{Sikivie:1983ip,DiLuzio:2020wdo}. Experiments such as ADMX~\cite{ADMX:2009iij,ADMX:2019uok}, CAPP~\cite{Kim:2023vpo}, UF~\cite{Hagmann:1996qd}, TASEH~\cite{TASEH:2022vvu}, HAYSTAC~\cite{HAYSTAC:2018rwy} and QUAX~\cite{QUAX:2023gop,QUAX:2024fut} were able to reach the QCD axion dark matter band already, with the projections of future experiments also documented in Ref.~\cite{AxionLimits} 

The second criteria comes from considering the axion production mechanism. The first of which is the misalignment mechanism~\cite{Sikivie:2006ni,Marsh:2015xka}. Such a process is expected because the axion field potential is essentially flat in the early Universe. Therefore, its field values will not correspond to the late time vacuum expectation value. The difference is typically described by the initial misalignment angle $\theta_i=a_i/f_a$, where the minimum has been chosen to be at zero. Due to the periodicity of the axion potential, $\theta_i$'s values are between $-\pi$ and $\pi$. At early times the field is trapped at $\theta_i$ due to Hubble friction but through cosmic expansion, it begins to oscillate around its true vacuum. The energy produced from these oscillations takes the form of cold matter. The relic abundance for the misalignment mechanism is given by
\begin{align}
    \Omega_a h^2\approx 0.08 \left(\frac{\theta_{\rm i}}{1}\right)^{2}\left(\frac{5.6~\mu {\rm eV}}{m_a}\right)^{7/6} &\text{ for } m_a \gtrsim 3\, H(T_{\rm QCD}), 
    \label{eq:relic_std}
\end{align}
where we only show the relation for the $m_a\gtrsim 3\, H(T_{\rm QCD})$ regime as it is what will be relevant in this paper. At the time of PQ breaking, each Hubble patch in the Universe will have random values of $\theta_i$. In the post-inflationary breaking scenario, the resultant $\theta_i$ is then averaged out by each Hubble patch coming into causal contact. This number is estimated to be $\langle\theta_i\rangle\approx\pi/\sqrt{3}$, however by including non-harmonic terms in the axion potential one obtains the numerical result $\theta_i\approx 2.15$~\cite{GrillidiCortona:2015jxo}. Nonetheless, $\theta_i$ should be some $\mathcal{O}(1)$ number, leading one to estimate a minimum value for $m_a$, which in turn leads to the upper bound on $f_a$ via eq.~(\ref{eq:QCDaxionmass}). 

We should comment here that typically the QCD axion/dark matter search extends to much higher values of the decay constant, the ultimate limit coming from isocurvature perturbations, $f_a\lesssim10^{16}\,{\rm GeV}$~\cite{Hertzberg:2008wr}. This is achieved by allowing $\theta_i$ to be tuned to a very small number $\theta_i\ll10^{-1}$. This can occur in pre-inflationary PQ breaking, but not the post-inflationary breaking scenario relevant for our work here. Furthermore, the axion abundance due to the misalignment mechanism can be diluted via some non-standard early cosmology~\cite{Steinhardt:1983ia, Lazarides:1990xp, Kawasaki:1995vt,Visinelli:2009kt,Nelson:2018via,Ramberg:2019dgi, Arias:2021rer,Arias:2022qjt, Xu:2023lxw}. However, these scenarios require additional new physics and they are therefore counter to the minimizing ethos of the preferred axion model framework. The only case, as far as the authors are aware, where no additional physics is introduced but early matter domination occurs is considered in Ref.~\cite{Cheek:2023fht}. 

The minimum value of $f_a$ comes from the results from numerical simulations of cosmic strings and their production of axions~\cite{Hagmann:2000ja,Wantz:2009it, Hiramatsu:2010yu, Kawasaki:2014sqa, Gorghetto:2020qws, Buschmann:2021sdq}. Recent developments from multiple independent collaborations~\cite{Gorghetto:2020qws,Buschmann:2021sdq,Saikawa:2024bta,Kim:2024wku} point to a minimum value of $f_a$, above which the correct relic will not be obtained. We follow Ref.~\cite{Alonso-Alvarez:2023wig} and take $5\times 10^9\, {\rm GeV}$ as the bound. We do this as an indicator of the region in parameter space when the cosmic string production of axions is able to reproduce to correct relic density.    

The third criteria for the preferred axion models comes from the realization that $\mathcal{L}_{\rm PQ}$ does not contain any term that allows for $Q$ decay. Stable $Q$ particles would cause numerous phenomenological problems in the post-inflationary breaking scenario. This is because at high temperatures, the strongly coupled $Q$'s will be in thermal equilibrium. Later at temperatures $T\lesssim m_Q$ the heavy quarks will undergo thermal freeze-out, and for $m_Q\gtrsim 10^6\, {\rm GeV}$ these particles will be overproduced~\cite{DeLuca:2018mzn}. There is the temptation to identify these particles with dark matter, but such a scenario is highly constrained~\cite{RICH1987173, Albuquerque:2003ei,Mack:2007xj,Kavanagh:2017cru, DeLuca:2018mzn}. Instead, one asserts that these new quarks must be unstable. To write decay terms for $Q$ one must assign specific PQ charges to the left and right-handed $Q$ fermions. Refs.~\cite{DiLuzio:2017pfr,Alonso-Alvarez:2023wig} enumerated the possible operators that lead to $Q$-decay for KSVZ-I; Depending on the Peccei-Quinn (PQ) charge assignment of the two chiral components of \(Q\), four distinct operators of dimension \(d \leq 4\) are possible:
\begin{equation}
\begin{aligned}
    \mathcal{O}^M_4 =& M_d \overline{Q}_L d_R, \quad &\text{for }& (\chi_L, \chi_R) = (0, -1), &\quad&\text{Model A},\\
    \mathcal{O}^H_4 =& y_{1,q} H \overline{q}_L Q_R, \quad&\text{for }& (\chi_L, \chi_R) = (1, 0),&\quad&\text{Model B}, \\
    \mathcal{O}^\Phi_4 =& y_{2,d} \Phi \overline{Q}_L d_R, \quad&\text{for }& (\chi_L, \chi_R) = (1, 0),&\quad&\text{Model B}, \\
    \mathcal{O}^{\Phi^\dagger}_4 =& y_{3,d} \Phi^\dagger \overline{Q}_L d_R, \quad &\text{for }& (\chi_L, \chi_R) = (-1, -2),&\quad&\text{Model C}. \\
\end{aligned}
\label{eq:dim4_models}
\end{equation}
Here, \(y_{n,d}\) represents Yukawa couplings of quarks with flavour \(d = d, s, b\), and $d_R$ and $q_L$ are the right-handed and left-handed SM fields, respectively. The $\left(\chi_L, \chi_R\right)$ denote the chiral PQ charges of the $Q$. Multiple operators can occur under the exact same charge assignment, we therefore group them as such, above we see three distinct models.  At dimension 5, additional operators are possible\footnote{The second term listed for model D is missing in~\cite{Alonso-Alvarez:2023wig}. We believe it has no effect on their conclusions. For our work, it is of consequence.}:
\begin{equation}
    \begin{aligned}
    \mathcal{O}^{|H|^2}_5 =& \frac{\lambda_d}{\Lambda} |H|^2 \overline{Q}_L d_R, &\text{for }& (\chi_L, \chi_R) = (0, -1), &\quad&\text{Model A}, \\
    \mathcal{O}^{|\Phi|^2}_5 =& \frac{\lambda'_d}{\Lambda} |\Phi|^2 \overline{Q}_L d_R, &\text{for }& (\chi_L, \chi_R) = (0, -1), &\quad&\text{Model A},\\
    \mathcal{O}^H_5 =& \frac{\lambda_{1,d}}{\Lambda} \Phi H \overline{d}_L Q_R, &\text{for }& (\chi_L, \chi_R) = (0, -1), &\quad&\text{Model A}, \\
    \mathcal{O}^\Phi_5 =& \frac{\lambda_{2,d}}{\Lambda} \Phi^2 \overline{Q}_L d_R, 
 &\text{for }& (\chi_L, \chi_R) = (2, 1), &\quad&\text{Model D}, \\
    \mathcal{O}^{\Phi H}_5 =& \frac{\lambda_{2,q}}{\Lambda} \overline{Q}_Rq_LH^{\dagger}\Phi, 
 &\text{for }& (\chi_L, \chi_R) = (2, 1), &\quad&\text{Model D},\\
    \mathcal{O}^{\Phi^\dagger}_5 =& \frac{\lambda_{3,d}}{\Lambda} (\Phi^\dagger)^2 \overline{Q}_L d_R, 
 &\text{for }& (\chi_L, \chi_R) = (-2, -3), &\quad&\text{Model E}. \\
\end{aligned}
\label{eq:dim5_models}
\end{equation}
Note that the KSVZ-II case is completely analogous under the exchanges \(u_R \leftrightarrow d_R\), \(H \leftrightarrow \tilde{H}\). Since the model is determined by the SM group and PQ charge assignments, there are only 10 preferred axion models. A priori there is no reason why the couplings of said operators should be substantially different. Hence, for the purposes of this paper we assume that for a given model all couplings are unity, i.e. for model B, $y_{1,d}= y_{2,d}=1$. 

The suppression scale, $\Lambda$ is typically assumed to be around the Planck scale, $m_{\rm Pl}=1.22\times10^{19}\,{\rm GeV}$, due to the conjecture that global symmetries must be broken at least by quantum gravity~\cite{Banks:2010zn, Reece:2023czb}. The global symmetry that is broken by the $Q$ decay terms is the $U(1)_{Q}$ symmetry which is present in eq.~(\ref{eq:LagPQ}). In principle $\Lambda$ could be much lower than $m_{\rm Pl}$, so taking $\Lambda=m_{\rm Pl }$ can be seen as simply taking the limiting case. Although there is an additional reason to consider $\Lambda=m_{\rm Pl }$. The $U(1)_{\rm PQ}$ symmetry must also be broken by the Planck scale, this famously has the potential to undermine solution to the strong CP problem and is known as the axion quality problem. It has been shown that effective operators that break PQ must have a dimension $d>11$~\cite{Kamionkowski:1992mf,Holman:1992us,Barr:1992qq}. If we were to lower this breaking scale ($\Lambda_{\rm PQ}$), the quality would need to increase (i.e., $d\gg11$), which would be undesirable. In order to minimize the number of free parameters $\Lambda=\Lambda_{\rm PQ}=m_{\rm Pl}$.

In principle, we could go further and look for models that only allow $Q$ decays via dimension $6$ or higher operators.  Refs.~\cite{DiLuzio:2016sbl,DiLuzio:2017pfr} used an approximate approach to estimate what was the highest dimension of decay that was still consistent with observations. For example, when the lifetime of the heavy quarks is so long, such that they disturb Big Bang Nucleosynthesis (BBN)~\cite{Kawasaki:2000en,Hannestad:2004px,Ichikawa:2005vw,Ichikawa:2006vm,deSalas:2015glj,Hasegawa:2019jsa}, or when the misalignment mechanism would overproduce axions. This leads to the condition that preferred axion models must allow heavy quarks to decay at most via dimension 5 effective operators. Recently however, Ref.~\cite{Cheek:2023fht} showed that the large abundance of $Q$ particles could alter the predictions of the misalignment abundance of axions, opening the possibility for higher dimension operators. More pertinent to the standard preferred axion models listed in eqs~(\ref{eq:dim4_models}) and (\ref{eq:dim5_models}), Ref.~\cite{Cheek:2023fht} found that models with dimension 5 decays could be long lived enough to provide a significant enhancement to the energy density of relic axions. These relic axions will contribute to the number of additional relativistic degrees of freedom, $\Delta N_{\rm eff}$. This contribution may be so large that measurements of the CMB may be used to constrain such models. In the next section we detail the decay widths for each of the preferred axion models. Note that after the publication of this article as a preprint, Ref.~\cite{DiLuzio:2024xnt} appeared, in which the authors find $d=6$ axion models that satisfy $N_{\rm DW}= 1$. Since $N_{\rm DW}$ depends only on the SM charges of $Q$, Ref.~\cite{DiLuzio:2024xnt} did not elaborate on the allowed PQ charges and resultant decay terms as in eqs.~(\ref{eq:dim4_models}) and (\ref{eq:dim5_models}). In section~\ref{sec:dNeff} we comment on our expectation for such models, but keep our primary focus on the $d\leq 5$ models described above.

\section{Calculating decay widths}
\label{sec:decay_width}
The decay terms in eqs.~(\ref{eq:dim4_models}) and~(\ref{eq:dim5_models}) contain two scalars, $\Phi$ and $H$, that get spontaneously broken at separate temperatures, $T_{\rm PQ}$ and $T_{\rm EW}$. In this work, $Q$ decay occurs in the $U(1)_{\rm PQ}$ broken phase so $T\lesssim T_{\rm PQ}\sim f_a$. Whereas for the $H$, the electroweak phase transition occurs at around $T_{\rm EW}\sim 100\, {\rm GeV}$ \cite{Kajantie:1996qd,DOnofrio:2012phz,DOnofrio:2014rug,DOnofrio:2015gop}, so decays could occur above or below this transition. In the unbroken Higgs phase we have straightforward decay terms such as $\mathcal{O}_4^H$, which leads to the $Q\to H\, q$ decay, with the decay width
\begin{equation}
    \Gamma\left(Q\to H\,q\right)=\sum_q\frac{y_{1,q}}{32\pi} m_Q =\frac{3}{32\pi} m_Q,\label{eq:modelB_Hq}
\end{equation}
where we sum over each family of SM quarks and take the massless limit for product particles. For the second equality we assume $\lambda_{1,q}=1$ for each family of quark. Additionally, at dimension 5, there exists the $\mathcal{O}_5^{\vert H\vert^2}$ term for model A. This results in three-body decays $Q\to H\,H^\dagger d$, leading to  
\begin{equation}
    \Gamma\left(Q\to d\,H\,H^\dagger\right)=\sum_{q=d,s,b}\frac{1}{768\pi^3}\left(\frac{\lambda_d}{\Lambda}\right)^2m_Q^3=\frac{1}{256\pi^3}\left(\frac{1}{\Lambda}\right)^2m_Q^3.
\end{equation}
We can estimate whether these decays will likely occur above or below the electroweak phase transition by solving for the temperature in the relation $3H_{\rm rad}(T_{\rm decay})\approx \Gamma$. Where $H_{\rm rad}$ is the Hubble rate due to radiation only, therefore, 
\begin{equation}
    T_{\rm decay}\approx\left(\frac{10}{g_\star(T_{\rm decay})}\right)^{1/4} \left(\frac{M_{\rm Pl} \Gamma_{Q}}{\pi}\right)^{1/2}.\label{eq:TdecHrad}
\end{equation}
Strictly speaking, if one allows for non-standard cosmologies, one should find the temperature using the exact expression of the Hubble parameter including the $Q$ matter contribution. This method only diverges from eq.~(\ref{eq:TdecHrad}) when $m_Q\gtrsim 10^{12}\,{\rm GeV}$ in our scenario (more details in section~\ref{sec:dNeff}), and even then only to a small degree. Therefore, for illustrative purposes we use the approximate form in eq.~(\ref{eq:TdecHrad}).

Taking $m_Q=10^{10}\,{\rm GeV}$, we see that the $\mathcal{O}_4^H$ decay implies $T_{\rm decay}\sim 10^{13}\,{\rm GeV}$ and the $\mathcal{O}_5^{\vert H\vert^2}$ decay implies $T_{\rm decay}\sim 10^{3}\,{\rm GeV}$. In this paper we will remain in the unbroken electroweak phase, performing consistency checks throughout. 
Returning to the broken $\Phi$, we can substitute eq.~(\ref{eq:phi_broke}) into the decay terms of the axion models. By expanding out the exponential term, we obtain $Q$ decay terms into SM and $a$ fields \footnote{We assume $m_\phi>2m_Q$ such that $\phi$ quickly decays to $Q$ particles and antiparticles before $Q$ freeze-out, if this condition were not satisfied, one would have to ensure the decay of $\phi$ somehow.}. Take the example of the model C term $\mathcal{O}_4^{\Phi^\dagger}$,
\begin{equation}
    \mathcal{O}_4^{\Phi^\dagger}=y_{3,d} \Phi^\dagger \overline{Q}_L d_R\to y_{3,d} \frac{1}{\sqrt{2}}f_a\left[1 -\frac{i a}{f_a}-\frac{ a^2}{2 f_a^2}+\dots\right] \overline{Q}_L d_R \label{eq:modelC_ad_aad}
\end{equation}
the first term is a mass term similar to $\mathcal{O}_4^M$ found in model A, which we will return to. The second and third terms give $Q$ decays into axions $Q\to a \,d$ and $Q\to a\,a\, d$, respectively. The decay widths for each are 
\begin{equation}
    \Gamma\left(Q\to a\, d\right)=\frac{y_{3,d}^2}{64\pi}m_Q\quad \textrm{and}\quad \Gamma\left(Q\to a\,a\, d\right)=\left(\frac{y_{3,d}^2}{f_a^2}\right)\frac{m_Q^3}{6144\pi^3}, 
\end{equation}
comparing these two widths 
\begin{equation}
    \frac{\Gamma\left(Q\to a\, d\right)}{\Gamma\left(Q\to a\,a\, d\right)}=96\pi^2\frac{1}{y_Q^2}\gtrsim 36\pi
\end{equation}
where we get the last inequality by imposing perturbative unitarity of the Yukawa coupling $y_Q\lesssim\sqrt{8\pi/3}$. With this result we see that $Q\to a\, d$ decay dominates over decays with multiple axions. Notice that this decay width is the same as for $Q\to H\,q$ in model B eq.~(\ref{eq:modelB_Hq}), hence models B and C have rapid decay rates and will not provide a period where thermally decoupled $Q$ particles free-stream. This is expected due to the renormalizability of the decay term.

On the other, model A has a renormalizable decay term, but fast decay does not actually occur. Let's show this in detail, the renormalizable term is
\begin{equation}
    \mathcal{O}_4^M = M_d \overline{Q}_Ld_R, 
    \label{eq:dim4mass}
\end{equation}
where $M_d$ can be different for each generation in principle. The additional term in \eqref{eq:dim4mass} then introduces an off-diagonal element in the mass matrix for $(d_i, Q)$\footnote{For KSVZ-II the mass matrix will be in the $(u_i, Q)$ basis.}:
\begin{align}
\begin{pmatrix}
\overline{d}_L &\overline{s}_L&\overline{b}_L  & \overline{Q}_L
\end{pmatrix}
\begin{pmatrix}
m_d & 0 & 0 & 0\\0 & m_s & 0 & 0\\0 & 0 & m_b & 0\\ M_d & M_s & M_b & m_Q
\end{pmatrix}
\begin{pmatrix}
d_R\\s_R\\b_R \\  Q_R
\end{pmatrix} 
+ {\rm h.c}.=
\begin{pmatrix}
\overline{d}_L & \overline{Q}_L
\end{pmatrix}
\begin{pmatrix}
m_d & 0 \\ M_d & m_Q
\end{pmatrix}
\begin{pmatrix}
d_R \\  Q_R
\end{pmatrix} 
+ {\rm h.c}~.
\end{align}
It should be understood that this is a $4\times4$ matrix which we represent in block-form on the right. The SM $3\times 3$ mass matrix $m_d$ is block diagonal. However, the above expressions are only relevant after electroweak symmetry breaking. Before the symmetry breaking, the SM quarks do not have their bare-mass, so $m_d\to 0$. For the same reason, the $\overline{d}_L Q_R$ operator does not exist. Therefore, in the unbroken Higgs phase we have
\begin{equation}
    \begin{pmatrix}
\overline{d}_L & \overline{Q}_L
\end{pmatrix}
\begin{pmatrix}
0 & 0 \\ M_d & m_Q
\end{pmatrix}
\begin{pmatrix}
d_R \\  Q_R
\end{pmatrix} 
+ {\rm h.c}~,
\end{equation}
in block form. We can perform a field re-definition of $(d,Q)$ and obtain the diagonalized mass matrix. It turns out that only a rotation in the right-handed sector is required,
\begin{equation}
\tan \theta_L  = 0 \quad \textrm{and} \quad
\tan \theta_R  = \frac{M_d}{m_Q}.
\label{eq:mixing_angle_full}
\end{equation}
Since the mixing between SM quarks and $Q$ is only in the right-handed sector, the kinetic terms do not lead to $Q$ decay terms. Decay through mixing only happens when $\theta_L\neq 0$ and this is necessarily occurs when SM quarks obtain masses through electroweak spontaneous symmetry breaking. We will show in the next section that for all models and parameters considered in this work, $Q$ decay will occur at temperatures above the electroweak phase transition $T_{\rm EW}\sim 100\, {\rm GeV}$. Therefore, Model A can only decay through its dimension 5 decay terms in eqs.~(\ref{eq:dim5_models}).

In the discussion above, we have already provided most of the information required to determine the decay widths for all dimension 5 decays. For example, taking model E and expanding the $\Phi$-field 
\begin{equation}
    \mathcal{O}^{\Phi^\dagger}_5 = \frac{\lambda_{3,d}}{\Lambda} (\Phi^\dagger)^2 \overline{Q}_L d_R=  \frac{\lambda_{3,d} f_a^2}{2\Lambda}  \overline{Q}_L d_R + \frac{i\,\lambda_{3,d} f_a}{\Lambda} a \overline{Q}_L d_R - \frac{\lambda_{3,d}}{\Lambda} a^2 \overline{Q}_L d_R+\dots
\end{equation}
the first term is simply the $\mathcal{O}_4^M$ operator but with $M_d\to \lambda_{3,d} f_a^2/2\Lambda$. The second and third terms are similar to those in Eq.~(\ref{eq:modelC_ad_aad}) such that the $Q\to a\,d$ and $Q\to a\,a\, d$ decays can be obtained by making the substitutions $y_{3,d}\to\sqrt{2}\lambda_{3,d}f_a/\Lambda$ and $y_{3,d}\to2\sqrt{2}\lambda_{3,d}f_a/\Lambda$, respectively. Similar associations can be made for the remaining decay terms in models A and D, with the exception of the $\mathcal{O}_5^{H}$ and $\mathcal{O}_5^{\Phi\,H}$ operators, giving rise to $Q\to q H^\dagger a$ decays. Due to the three unique product particles, the resulting decay width is half of that of $Q\to a\,a\,d$ 
\begin{equation}
    \Gamma\left(Q\to H\,q_L \, a\right)=\frac{1}{512\pi^3}\left(\frac{\lambda}{\Lambda}\right)^2m_Q^3.
\end{equation}

In table~\ref{tab:decay_summary} we summarise this discussion and present all the decay rates in a compact way. We have highlighted the dominant decay terms for each model under the relevant assumptions for this paper, i.e. that $m_Q=f_a\gtrsim 5\times 10^9\,{\rm GeV}$. Table~\ref{tab:decay_summary} shows that only model A has $Q\to {\rm SM}$ decay channel dominating, and for the others, there is the possibility that axion production through $Q$ decay contributes to the cosmic energy budget as dark radiation. This contribution depends on the axion-SM interaction rate and whether, at the time of $Q$ decay, the axion is thermally decoupled and whether it remains decoupled. We discuss this in the next section. 

\begin{table}
    \centering
    \begin{tabular}{c|c|c|c|c|c}
        Decay & Model A & Model B &  Model C & Model D & Model E \\
        \hline
         $\Gamma(Q\to a\, d) = \displaystyle{\frac{3}{32\pi}}\mathcal{C}^2 m_Q$& \rule[0.5ex]{2em}{0.5pt} & $\cellcolor{lightblue}\displaystyle{\frac{y_{2,d}}{\sqrt{2}}}$ & $\cellcolor{lightblue}\displaystyle{\frac{y_{3,d}}{\sqrt{2}}}$ & \cellcolor{lightblue}$\displaystyle{\frac{\lambda_{2,d}f_a}{\Lambda}}$ & $\cellcolor{lightblue}\displaystyle{\frac{\lambda_{3,d}f_a}{\Lambda}}$\\
         $\Gamma(Q\to H\, q_L) =\displaystyle{\frac{3}{32\pi}}\mathcal{C}^2 m_Q$& $\cellcolor{lightblue}\displaystyle{\frac{\lambda_{1,q}f_a}{\Lambda}}$ & \cellcolor{lightblue}$y_{1,q}$ & \rule[0.5ex]{2em}{0.5pt} & \cellcolor{lightblue}$\displaystyle{\frac{\lambda_{2,q}f_a}{\Lambda}}$ & \rule[0.5ex]{2em}{0.5pt} \\
        $\Gamma(Q\to a\,a\,\, d) = \displaystyle{\frac{1}{256\pi^3}}\mathcal{C}^2 m_Q^3$ & \rule[0.5ex]{2em}{0.5pt} & $\displaystyle{\frac{y_{2,d}}{2f_a}} $ & $\displaystyle{\frac{y_{3,d}}{2f_a}}  $& \rule[0.5ex]{2em}{0.5pt} & $\displaystyle{\frac{\lambda_{3,d}}{\Lambda}}$\\
         $\Gamma(Q\to H\,H\,\, d) = \displaystyle{\frac{1}{256\pi^3}}\mathcal{C}^2 m_Q^3$& $\displaystyle{\frac{\lambda_{d}}{\Lambda}}$ & \rule[0.5ex]{2em}{0.5pt} & \rule[0.5ex]{2em}{0.5pt} & \rule[0.5ex]{2em}{0.5pt} & \rule[0.5ex]{2em}{0.5pt}\\
         $\Gamma(Q\to H\,q_L\,\, a) = \displaystyle{\frac{1}{512\pi^3}}\mathcal{C}^2 m_Q^3$& $\displaystyle{\frac{\lambda_{1,q}}{\Lambda}}$ & \rule[0.5ex]{2em}{0.5pt} & \rule[0.5ex]{2em}{0.5pt} & $\displaystyle{\frac{\lambda_{2,q}}{\Lambda}}$ & \rule[0.5ex]{2em}{0.5pt}\\
         $M_d$   & $M_d$ & $\displaystyle{\frac{y_{2,d}f_a}{\sqrt{2}}}$ & $\displaystyle{\frac{y_{3,d}f_a}{\sqrt{2}}}$ &  $\displaystyle{\frac{\lambda_{2,d}f_a^2}{2\Lambda}}$& $\displaystyle{\frac{\lambda_{3,d}f_a^2}{2\Lambda}}$\\
    \end{tabular}
    \caption{Table summarizing the decay channels where each entry shows corresponding $\mathcal{C}$ expression that should be substituted in the $\Gamma$ for each row. For each model we shade in the decay channel which dominates when $m_Q=f_a\gtrsim 5\times 10^{9}\,{\rm GeV}$, as is the focus of this study. For models B and D, two channels contribute approximately equal amounts. We have kept the coefficients as in Eqs.~(\ref{eq:dim4_models}) and (\ref{eq:dim5_models}), but assumed that they are family universal, so multipled by $3$ when appropriate. The bottom row shows the equivalent mass mixing term $M_d$ as it occurs for $\mathcal{O}_4^M$, this term only contributes to the $Q$ decay at scales below the electroweak phase transition.  }
    \label{tab:decay_summary}
\end{table}

\section{Predictions of $\Delta N_{\rm eff}$}
\label{sec:dNeff}
The radiation energy density at late times, i.e. recombination is composed of standard model radiation and dark radiation, which in our case will be the axion,
\begin{equation}
    \rho_{\rm R} = \rho_{\rm R}^{\rm SM} + \rho_{a}
\end{equation}
we can relate this to the number of relativistic degrees of freedom $N_{\rm eff}^{\rm SM}$ and additional degrees of freedom $\Delta N_{\rm eff}$ via
\begin{equation}
\rho_\mathrm{R}\equiv \rho_\gamma\left[1+\frac{7}{8}\left(\frac{T_\nu}{T_\gamma}\right)^4(N_\mathrm{eff}^{\rm SM}+\Delta N_\mathrm{eff})\right]\,, 
\end{equation}
where the reported theoretical value $N_{\rm eff}^{\rm SM} = 3.044$ coming from the SM neutrinos is calculated in Refs.~\cite{Mangano:2001iu,deSalas:2016ztq,Gariazzo:2019gyi,Akita:2020szl,Froustey:2020mcq,Bennett:2020zkv, Drewes:2024wbw}. Solving for $\Delta N_{\rm eff}$
\begin{equation}
    \Delta N_{\rm eff} \equiv \left\{\frac{8}{7}\left(\frac{4}{11}\right)^{-\frac{4}{3}}+N_{\rm eff}^{\rm SM}\right\} 
 \frac{\rho_{a}}{\rho_{\rm R}^{\rm SM}}\,.
 \label{eq:DNEFF_eq}
\end{equation}
Both $\rho_{\rm R}^{\rm SM}$ and $\rho_{a}$ are evaluated at the time relevant to the experimental measurement one is considering. We will compare our predictions with measurements of the CMB, so  matter-radiation equality is the relevant time, $T_{\rm eq}=0.75$ eV. As estimated in the previous section, some preferred axion models predict $Q$ decay some time after the heavy quarks have frozen out. Exploiting the free-streaming of matter $\rho_m\propto a^{-3}$ vs. radiation $\rho_r\propto a^{-4}$ it is reasonable to entertain the possibility that a very large $Q$ particle can come to dominate the early Universe. As was estimated in~\cite{Cheek:2023fht}, for $m_Q$ values between $10^{10}-10^{12}\,{\rm GeV}$, dimension 5 decays typically occur at the interface between radiation domination and early matter domination. A late-time $Q$ decay will, depending on the model, produce boosted axions with energy $\sim m_Q/2$. If at the time of decay, axions have thermally decoupled from the SM plasma, the Boltzmann equations are 
\begin{subequations}\label{eq:FBEqs}
\bea
\frac{\dd s_\mathrm{R}^{\rm SM}}{\dd t} +3H s_\mathrm{R}^{\rm SM}&=&   \frac{{\rm BR}_{\rm SM}\Gamma_{Q}}{T}\rho_{Q} \,,\\
\frac{\dd\rho_a}{\dd t} +  4H \rho_a &=& {\rm BR}_{a}\Gamma_{Q}\rho_{Q}\,,\label{eq:decoupled_a_BE}\\
 \frac{\dd \rho_{Q}}{\dd t}+3H \rho_{Q} &=&  -\Gamma_{Q} n_Q  \label{eq:decoupled_Q_BE}\,, 
\eea\label{eq:decoupled_BEs}
\end{subequations}
where $s_{R}^{\rm SM}$ is the SM entropy density, $n_Q$ is the number density of $Q$, ${\rm BR}_{a/{\rm SM}}$ is a generalized branching ratio that denotes the share of energy from the $Q$ decay that goes into the axion or SM, respectively. For example, if only $Q\to a\, d$ decay is present, then ${\rm BR}_a$ is $1/2$ because half the energy is going into the axion sector. Since $Q$'s can only decay into SM or $a$, ${\rm BR}_{\rm SM} + {\rm BR}_{a}=1$. The Hubble parameter above, $H$, is determined by Friedmann equation 
\be\label{eq:Hubble}
\frac{3H^2M_{\rm Pl}^2}{8\pi}=\rho_\mathrm{R}^{\rm SM} + \rho_{a} + \rho_{Q}\,,
\ee
In order to ensure that the decoupled equations above are valid, we need to check that the axions have thermally decoupled at earlier times and do not rethermalize at time of $Q$ decay. 

\subsection{Thermal axion decoupling}
The thermal behavior of the axion in the early Universe is governed by the Boltzmann equation describing the scattering of the axion 
\begin{equation}
    \frac{{\rm d} n_a}{{\rm d} t} + 3 H n_a = \gamma_a\left(1- \frac{n_a}{n_a^{\rm eq}}\right)\label{eq:axion_thermal_BE}
\end{equation}
where $\gamma_a$ is the thermal axion production rate~\cite{Graf:2010tv,Salvio:2013iaa,DEramo:2021lgb,Bouzoud:2024bom} defined as 
\begin{equation}
    \gamma_a=\sum_{i,j} n_i^{\rm eq} n_j^{\rm eq}\langle\sigma v\rangle_{i\,j\leftrightarrow k\,a}, \label{eq:gamma_sum}
\end{equation}
where particles, $i,\, j$ and $k$ are in thermal equilibrium. Using the non-linear KSVZ Lagrangian 
\begin{equation}
    \mathcal{L}_{\text{KSVZ}}^{\text{(non-linear)}} = \frac{1}{2} \partial^\mu a \, \partial_\mu a + \overline{Q} i \slashed{D} Q - \frac{\partial_\mu a}{2 v_\varphi} \overline{Q} \gamma^\mu \gamma^5 Q + \frac{\alpha_s}{8 \pi} \frac{a}{f_a} G^A_{\mu \nu} \widetilde{G}^{A \mu \nu},
\end{equation}
one finds that at high temperatures $T\gtrsim m_Q$, $Q$ annihilation and scattering drives the axion production rate, $Q+\overline{Q} \to g + a$ and $Q/\overline{Q} + g \to  Q/\overline{Q} + a$. Performing the thermal averaging as in Ref.~\cite{DEramo:2021lgb} one obtains a $\gamma_a\sim T^4 e^{-m_Q/T}$ dependence. 

At temperatures below $m_Q$, axion production is driven by quark and gluon scatterings. These interactions exhibit IR divergences, therefore calculating $\gamma_a$ via the standard methods used in dark matter physics~\cite{Gondolo:1990dk} is non-trivial. Instead, the gluon/quark scattering contribution to $\gamma_a$ can be calculated by using thermal field theory~\cite{Braaten:1991jj, Graf:2010tv}, we take the results of Refs.~\cite{Salvio:2013iaa, DEramo:2021lgb} 
\begin{equation}
    \gamma_{gg}  = \frac{16 \, \zeta(3) \, }{\pi^3} \left( \frac{\tilde{c}_g^{Q}(T) \, \alpha_s}{8 \pi \, f_a} \right)^2 F_3(T) \, T^6,
\end{equation}
where $\tilde{c}_g^{Q}(T)$ is the effective gluon anomaly coefficient calculated from the 1PI effective action. In the low temperature limit $T\ll m_Q$, $\tilde{c}_g^{Q}(T)\to1$ and $T>m_Q$ the coefficient suppresses the rate $\sim T^{-2}$. The function $F_3(T)$ is the ``thermal function'' that parametrizes additional thermal effects coming from thermal corrections calculated in Refs.~\cite{Salvio:2013iaa,DEramo:2021lgb}\footnote{Note that recently Ref.~\cite{Bouzoud:2024bom} has performed an independent calculation of $\gamma_gg$ were the reported determination of $F_3$ is a factor of $2$ lower at $T=10\,  {\rm TeV}$ than Ref.~\cite{Salvio:2013iaa}. The final results in this paper are insensitive to this variation in $F_3$.}. Note that this function is only mildly dependent on $T$ and can be taken as constant for temperatures above the electroweak scale (see figure 13 of Ref.~\cite{DEramo:2021lgb}). Therefore, in the temperature regime $T<m_Q$ the axion production rate is $\gamma_a\propto T^6/f^2_a$.

\begin{figure}
    \centering
    \includegraphics[width=0.9\linewidth]{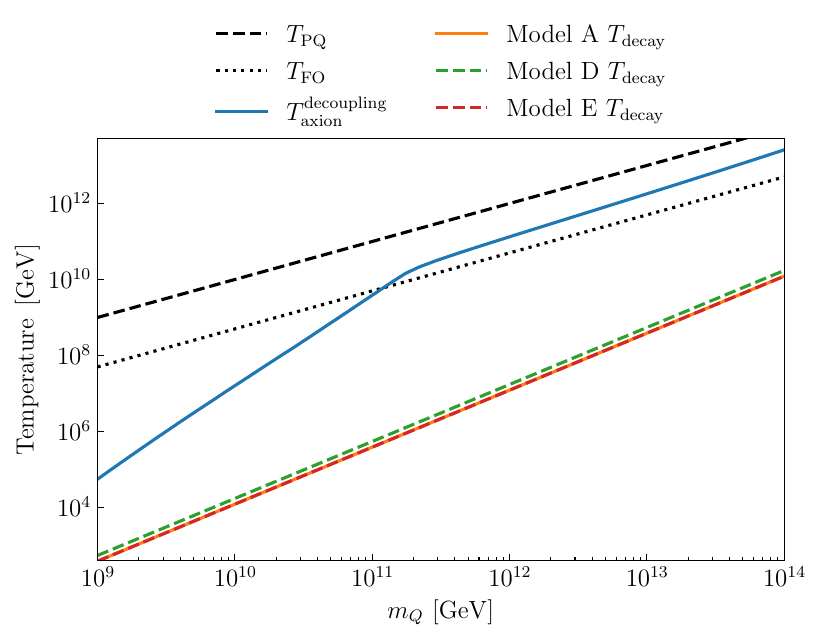}
    \caption{Showing the comparison in temperatures of the thermal bath for multiple events: PQ symmetry breaking ($T_{\rm PQ}\sim f_a$, black dashed); $Q$ thermal freeze-out ($T_{\rm FO}\sim m_Q/20$, black dotted line); thermal decoupling of the axion $T_{\rm ax}^{\rm d}$; and the decay of $Q$ particles for Model A (orange solid line), Model D (dashed green) and Model E (dashed red), respectively. We have set $m_Q=f_a$.}
    \label{fig:axion_decoupling}
\end{figure}

Returning back to the Boltzmann equation Eq.~(\ref{eq:axion_thermal_BE}), we can approximate when the axion scattering rate $\gamma_a$ will no longer be large enough to keep the axion in thermal equilibrium with the SM. Specifically we solve for temperature in the equality $3H(T)n_a(T)=\gamma_a(T)$ numerically. The results of this exercise are shown in figure~\ref{fig:axion_decoupling} with a blue line. We see there are two distinct $m_Q$ dependencies on the axion decoupling temperature. This corresponds to the dominant scattering at the moment when decoupling occurs. For high $m_Q$ the decoupling occurs when $Q$ annihilation and scattering is dominant, whereas the below $\sim 10^{11}\,{\rm GeV}$ decoupling occurs when gluon and quark scattering with axions is dominant.  

Furthermore, in figure~\ref{fig:axion_decoupling} we have plotted the approximate PQ symmetry breaking temperature $T_{\rm PQ}=f_a$ and the $Q$ freeze-out temperature $T_{\rm FO}\sim m_Q/20$. We do this to indicate the consistency of the post-inflationary PQ breaking picture. The reheating temperature after inflation must be above $T_{\rm PQ}$, therefore the figure shows that the axion will be in thermal equilibrium at early times and will decouple at a similar time to $Q$ freeze-out for high $m_Q$ values but substantially later for masses below $\sim 10^{11}\,{\rm GeV}$. 

Finally, we include the temperature of the SM bath at time of $Q$ decays for models A, D and E, the models that have slow decays. We see that in the case we considered here there are between 2 and 4 orders of magnitude difference between the temperature of decoupling and decay. This means the axions produced by $Q$ decay will be decoupled from the SM and take the form of dark radiation. That is of course as long as the boosted axion interaction rate with SM particles isn't enhanced. To formally prove that this is the case would require a recalculation of $\gamma_{gg}$ but now taking into account the non-thermal boosted axion distribution. This is beyond the scope of this work so we instead provide a physical argument. 

If we use the definition of $\gamma_a$ in eq.~(\ref{eq:gamma_sum}) and compare with the result given for $\gamma_{gg}$, we can infer that the thermal cross-section for gluon and quark scatterings with axions, $\langle \sigma v\rangle_{gg}$, is approximately independent of the SM temperature. By comparing the axion annihilation term in eq.~(\ref{eq:axion_thermal_BE}) to the Hubble term, we can define a ratio
\begin{equation}
    R(T)=\frac{3Hn_a n_a^{\rm eq}}{\gamma_a n_a}=\frac{3H} {n_i^{\rm eq}\langle \sigma v\rangle_{gg}}= \frac{T_{\rm ax}^{\rm d}}{T}
\end{equation}
where we've used the fact that $n_a^{\rm eq}=n_j^{\rm eq}$, i.e. assumed that the particle $j$ is a boson. We have also used the fact that $R(T_{\rm ax}^{\rm d})=1$ by definition. Therefore, as long a $T<T_{\rm ax}^{\rm d}$ then $R>1$ which corresponds to the decoupled axion regime. As shown in figure~\ref{fig:axion_decoupling}, for models A, D and E, the temperature at time of $Q$ decay is at least two orders of magnitude smaller than $T_{\rm ax}^{\rm d}$.

A more rigorous formulation of the above argument would require one to solve the Boltzmann equations for axion scattering on the level of the phase-space. This has been recently performed in Refs.~\cite{Bouzoud:2024bom,Badziak:2024qjg,DEramo:2024jhn} without the additional feature of a distribution of boosted axion particles. We leave this for future work. 

\subsection{Solving the Boltzmann equations}
In order to track the number densities of $Q$ and axions we split the early Universe dynamics into two periods. First, we solve for the thermal freeze-out of the heavy quarks
\begin{equation}
   \frac{\dd n_{Q}}{\dd t} + 3H n_{Q} =\langle\sigma\,v\rangle\left[ (n_Q^{\rm eq})^2-n_Q^2\right]  -\Gamma_{Q} n_Q \,,  
\end{equation}
where $\Gamma_Q$ is the total decay width of $Q$, which for models A, D and E, is small enough to have a negligible effect. The $Q/\bar{Q}$ annihilation is driven by the strong interaction, leading to the thermal cross-section 
\begin{equation}
    \langle\sigma v\rangle_{\bar{Q}Q}=\frac{\pi\alpha_s^2}{16 m_Q^2}\left(c_f n_f + c_g\right). \label{eq:FO_BE}
\end{equation}
where $n_f$ is the number of quark flavours that $Q$ can annihilate into, and $(c_f, c_g) = \left( 2/9, 220/27 \right)$ for colour triplets such as $Q$~\cite{Zyla:2020zbs}. In practice we rewrite this equation in terms of the yield of $Q$, $Y_Q=n_Q/s_{\rm R}$, where $s_{\rm R}$ is the entropy density and solve numerically to obtain the yield after freeze-out, $Y_{Q}(T < m_Q/50)$. 

Once the relic value of $Y_{Q}$ is determined we evolve the energy densities of thermal radiation and decoupled $Q$'s until the axion decoupling temperature $T_{\rm ax}^{\rm d}$ according to Eqs.~(\ref{eq:decoupled_BEs}) but since the axion is still in equilibrium we only evolve $\rho_Q$ and $s_{\rm R}$ because the axion abundance can be accounted for by including its contribution to number of entropy degrees of freedom, $g_{\star s}(T)$ in 
\begin{equation}
    s_{\rm R}=\frac{2\pi^2}{45}g_{\star s}(T)\,T^3.
\end{equation}
After $T_{\rm ax}^{\rm d}$ we then evolve all three equations in eqs~(\ref{eq:decoupled_BEs}) now with the initial axion energy given by the equilibrium value $\pi^2/30 \times T_i^4$. After this point both the axions and SM radiation receive an energy injection via $Q$ decay, but by differing amounts according to the dominant decay channels. From table~\ref{tab:decay_summary} we can see that model E will produce axions and SM radiation in roughly equal parts, whereas $Q$ decays from model A will only heat the SM radiation, decreasing $\rho_a/\rho_{\rm R}^{\rm SM}$ and therefore $\Delta N_{\rm eff}$~(\ref{eq:DNEFF_eq}).

Note that as is apparent in figure~\ref{fig:axion_decoupling}, for some values of $m_Q$, decoupling happens during the freeze-out period. In this case we calculate the axion energy density by 
\begin{equation}
    \rho_a= \frac{\pi^2}{30} (T_{\rm ax}^{\rm d})^4\left(\frac{g_\star(T)}{g_\star(T_{\rm ax}^{\rm d})}\right)^{4/3}\left(\frac{T}{T_{\rm ax}^{\rm d}}\right)^4\label{eq:free_stream_ax}
\end{equation}
and evolve the Boltzmann equations eqs~(\ref{eq:decoupled_BEs}). For models with fast decays, i.e. models B and C, one can use eq.~(\ref{eq:free_stream_ax}) directly to calculate $\Delta N_{\rm eff}$. Therefore, when $m_Q=f_a$ the estimated $\Delta N_{\rm eff}\approx 0.027$~\cite{Brust:2013ova,Baumann:2016wac} which is the standard prediction for a thermal axion that has decoupled above the electroweak scale.

\subsection{Results and discussion}

\begin{figure}
    \centering
    \includegraphics[width=0.9\linewidth]{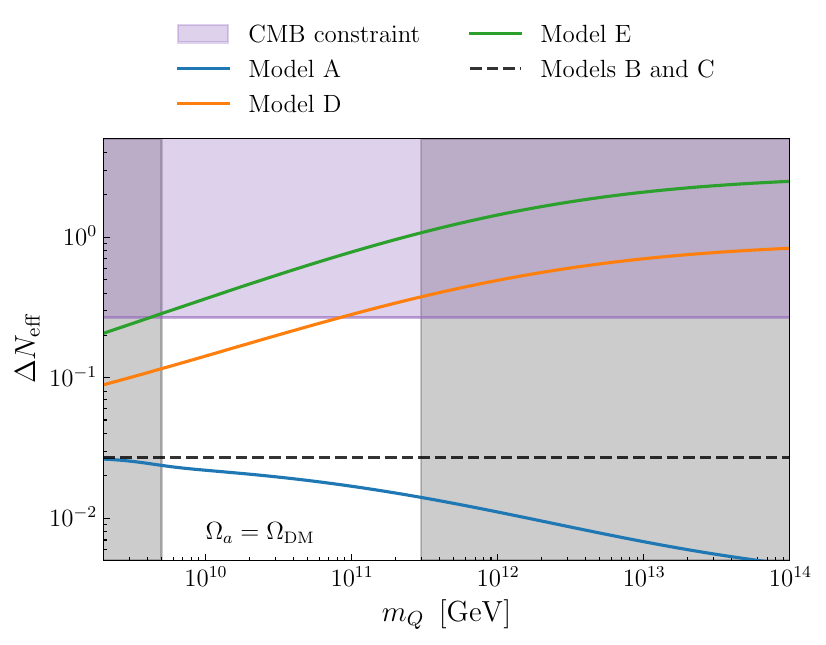}
    \caption{Predicted $\Delta N_{\rm eff}$ values for preferred axion models and different values of $m_Q$ which here is associated with the axion decay constant, $m_Q=f_a$. Models A, D and E are denoted with blue, orange and green solid lines respectively. The ``thermal axion'' prediction for $\Delta N_{\rm eff}$ is the same for models B and C, which we show with a black dashed line, see text for details. We show the current limits on $\Delta N_{\rm eff}$ as determined from measurements of the CMB from Planck~\cite{Planck:2018vyg} by way of a purple shaded region. The grey regions are disfavored by the requirement that the axion is dark matter, see criteria 2 in section~\ref{sec:pref_ax}. }
    \label{fig:main_results_DNEFF}
\end{figure}

In figure~\ref{fig:main_results_DNEFF} we show our results for the preferred axion models listed in eqs~(\ref{eq:dim4_models}) and (\ref{eq:dim5_models}) where we set $m_Q=f_a$. We compare our predictions to the current CMB limits exclude values of $\Delta N_{\rm eff}$. Specifically Ref.~\cite{Planck:2018vyg} reports $(N_{\rm eff})_{\rm P18}=2.88^{+0.44}_{-0.42}$, hence we constrain $\Delta N_{\rm eff}=(N_{\rm eff})_{\rm P18}-(N_{\rm eff})_{\rm SM}\leq 0.276$.  
Figure~\ref{fig:main_results_DNEFF} shows that there is a sizable contribution to $\Delta N_{\rm eff}$ from $Q$ decay for certain preferred axion models. As implied by the results in table~\ref{tab:decay_summary}, this is greatest for model E, which has $Q\to a\, d$ decays dominate. Given the additional $Q\to H q_L$ channel in model D the energy transfer to axions is half of that than in model E, as a result the $\Delta N_{\rm eff}$ values for model D are suppressed by $\sim 2$. Interestingly, our results show that model E is all but excluded by Planck for $m_Q=f_a$. 

Furthermore, model A provides a relative suppression in $\Delta N_{\rm eff}$, this is due to $Q$ decays heating SM particles but not the decoupled axions. The suppression being relative to the thermal axion value $\Delta N_{\rm eff}\approx 0.027$, which current experiments are insensitive to, means model A is currently indistinguishable to models B and C. Future measurements of the CMB have the objective of reach the sensitivities required to test the thermal axion scenario. These include CMB-HD~\cite{Sehgal:2019ewc,CMB-HD:2022bsz}, LiteBIRD~\cite{Matsumura:2013aja,LiteBIRD:2022cnt}, and PRISM~\cite{PRISM:2013fvg}. With these experiments in mind, it seems plausible that model A could be distinguished against models B and C. 

\begin{figure}
    \centering
    \includegraphics[width=0.9\linewidth]{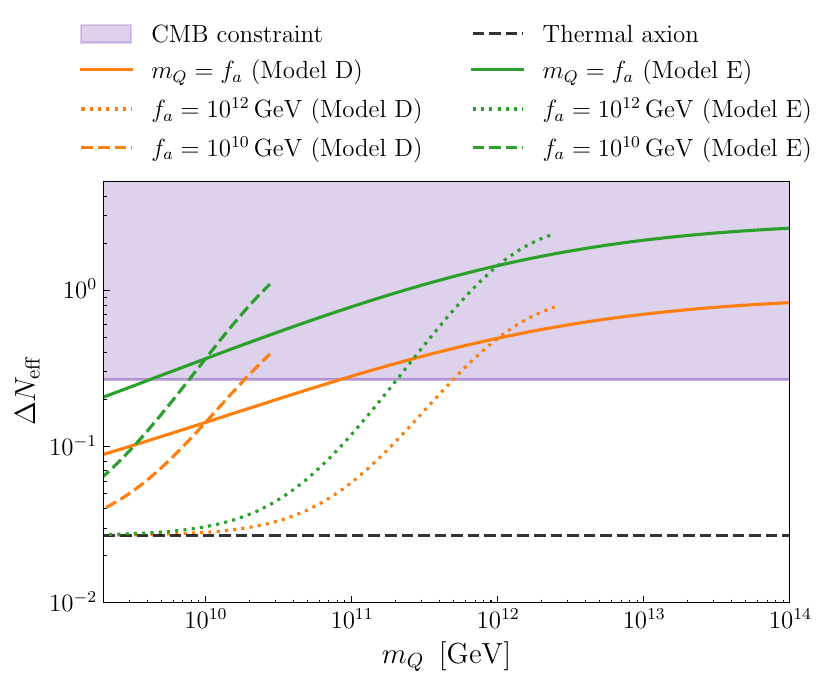}
    \caption{Same as figure~\ref{fig:main_results_DNEFF} but with only models D and E. We add the dotted and dashed lines to show how fixing $f_a$ while varying $m_Q$ alters the predicted $\Delta N_{\rm eff}$. We plot $m_Q$ values up to the perturbative unitarity limit of $y_Q$.}
    \label{fig:DNEFF_mQchange}
\end{figure}

Perhaps the current CMB limit shown in figure~\ref{fig:main_results_DNEFF} is reason enough to strip the title of \textit{preferred axion model} from model E, given that now model E would only avoid existing constraints if $m_Q<f_a$. This case could be made for model D as well if constraints on $\Delta N_{\rm eff}$ are improved in line with expectations. In figure~\ref{fig:DNEFF_mQchange} we show how $\Delta N_{\rm eff}$ values change when one allows $m_Q<f_a$. We exhibit this by showing $\Delta N_{\rm eff}$ for different $m_Q$ values but now fixing $f_a$, dashed colored lines are for $f_a=10^{10}\,{\rm GeV}$ and dotted for $f_a=10^{12}\,{\rm GeV}$. The maximum $m_Q$ is then determined by the perturbative unitarity limit on the Yukawa coupling $y_Q$. We see that for model E and $f_a=10^{12}\,{\rm GeV}$ roughly an order of magnitude of the allowed $m_Q$ values are still excluded. Indeed, model E only becomes indistinguishable from models B and C at around $m_Q\sim 10^{10}\,{\rm GeV}$. 

Finally we comment on possible axion models with effective decay terms $d\geq6$. The recent result of Ref.~\cite{DiLuzio:2024xnt} indicate that there exists four $d=6$ models that survive all cosmological tests including the $N_{\rm DW}=1$ criteria. Unfortunately the authors only provide `example operators' for the decay processes so we are unable to comment on the expected value for $\Delta N_{\rm eff}$ with complete confidence. However, by exploring the possible decays with the SM charge assignments of the new models, we believe that $Q$ decays will only have SM products, i.e. ${\rm BR}_a=0$. The result is that any $Q$ decays will dilute the thermal relic axion density, reducing the predicted $\Delta N_{\rm eff}$. This is similar to model A above but the suppression will be larger because dimension-6 decay will occur much later than for model A and the heavy quarks would have dominated for many e-folds, such that $\Delta N_{\rm eff}\ll 0.027$. If some models are found to have ${\rm BR}_a\neq 0$ then measurements of $\Delta N_{\rm eff}$ could provide constraints. As shown in Ref~\cite{Cheek:2023fht} this is currently set to be ${\rm BR}_a\lesssim 0.03$.

\section{Conclusion}
\label{sec:conclusion}
Axion dark matter is an attractive solution to the dark matter and strong CP problems. The preferred axion models were previously proposed as a set of axion models that have minimal new particle content and avoid cosmological problems such as domain walls. The resulting models contain a heavy quark, $Q$, which is required to decay into the SM particles and/or axions.

In this paper we have investigated the potential to use measurements of the number of relativistic degrees of freedom, $N_{\rm eff}$, to constrain these preferred axion models. It is usually assumed that these models do not suffer from such constraints because $Q$ decays before Big Bang Nucleosynthesis. However, as we have shown in this work, for many such models the decay rate and abundance of $Q$ particles can be significant enough to alter the expected axion contribution to the energy budget of the Universe. 

Our result is in contrast with the standard expectation for the thermal axion scenario. We find that 40\% of preferred axion models can enhance the value for $N_{\rm eff}$ to such an extent that Planck results already exclude regions of parameter space, see figure~\ref{fig:main_results_DNEFF}. With future experiments set to improve their sensitivity to $\Delta N_{\rm eff}$ by roughly an order of magnitude, phenomenologically distinguishing between preferred axion models is actually a possibility. 

\acknowledgments
We thank Shao-Feng Ge, Luca Visinelli, Matthew Kirk, Yu-Cheng Qiu, Jiang Zhu and Yong Du for their valuable discussions. The authors are supported by the National Natural Science Foundation of China (12425506, 12375101, 12090060 and 12090064) and the SJTU Double First Class start-up fund (WF220442604).

\bibliographystyle{JHEP}
\bibliography{ref}

\end{document}